\documentclass[journal=jacsat,manuscript=article]{achemso}

\usepackage[version=3]{mhchem}
\usepackage{physics}
\usepackage{diagbox}
\usepackage{subcaption}
\usepackage{xcolor}
\usepackage{natbib}
\usepackage{float}
\usepackage{etoolbox}
\makeatletter
\renewcommand*{\acs@author@fnsymbol@symbol}[1]{
    \ifcase #1 *\or
    1\or
    2\or
    3\or
    4\or
    5\or
    6\or
    7\or
    8\or
    9\or
    10
    \fi
}
\renewcommand*\acs@contact@details{
    {\sffamily *\,e-mail: \acs@email@list }%
    \acs@number@list
}          
\patchcmd{\acs@address@list@auxii}
{\acs@author@fnsymbol{\acs@affil@marker@cnt}}
{\textsuperscript{\acs@author@fnsymbol{\acs@affil@marker@cnt}}}
{}{}
\newcommand{\bea}{\begin{eqnarray*}}
	\newcommand{\eea}{\end{eqnarray*}}
\newcommand{\bne}{\begin{equation*}}
\newcommand{\ede}{\end{equation*}}

\newcommand{\bnen}{\begin{equation}}
\newcommand{\eden}{\end{equation}}
\newcommand{\bean}{\begin{eqnarray}}
\newcommand{\eean}{\end{eqnarray}}
\newcommand{\bsen}{\begin{subequations}}
	\newcommand{\esen}{\end{subequations}}

\newcommand{\bna}{\begin{array}}
	\newcommand{\eda}{\end{array}}
\newcommand{\bnm}{\begin{enumerate}}
	\newcommand{\edm}{\end{enumerate}}

\author{Abhikbrata Sarkar}
\email{abhikbrata.sarkar@student.unsw.edu.au}
\affiliation{School of Physics, The University of New South Wales, Sydney 2052, Australia}
\alsoaffiliation{Australian Research Council Centre of Excellence in Future Low-Energy Electronics Technologies, The University of New South Wales, Sydney NSW 2052, Australia}

\author{Pratik Chowdhury}
\affiliation{School of Physics, The University of New South Wales, Sydney 2052, Australia}

\author{Xuedong Hu}
\affiliation{Department of Physics, University at Buffalo, SUNY, Buffalo, NY 14260-1500}

\author{Andre Saraiva}
\affiliation{School of Electrical Engineering and Telecommunications, The University of New South Wales, Sydney 2052, Australia}

\author{A. S. Dzurak}
\affiliation{School of Electrical Engineering and Telecommunications, The University of New South Wales, Sydney 2052, Australia}

\author{A. R. Hamilton}
\affiliation{School of Physics, The University of New South Wales, Sydney 2052, Australia}
\alsoaffiliation{Australian Research Council Centre of Excellence in Future Low-Energy Electronics Technologies, The University of New South Wales, Sydney NSW 2052, Australia}

\author{Rajib Rahman}
\affiliation{School of Physics, The University of New South Wales, Sydney 2052, Australia}

\author{Dimitrie Culcer}
\email{d.culcer@unsw.edu.au}
\affiliation{School of Physics, The University of New South Wales, Sydney 2052, Australia}
\alsoaffiliation{Australian Research Council Centre of Excellence in Future Low-Energy Electronics Technologies, The University of New South Wales, Sydney NSW 2052, Australia}

\title[\texttt{achemso} demonstration]
{Effect of disorder and strain on the operation of planar Ge hole spin qubits}
\begin{document}

\begin{abstract}
Germanium quantum dots in strained $\text{Ge}/\text{Si}_{1-x}\text{Ge}_{x}$ heterostructures exhibit fast and coherent hole qubit control in experiments. Nevertheless, a full theoretical understanding of their underlying physics is lacking due to the absence of a systematic method of accounting for inhomogeneous strain, whose effects often overwhelm the bare spin-orbit coupling in the Luttinger Hamiltonian. Here we seek to remedy this shortcoming by theoretically and numerically addressing the effects of random alloy disorder and gate-induced strain on the operation of planar Ge hole spin qubits. We use the atomistic valence force field (VFF) method to compute the strain due to random alloy disorder, and thermal expansion models in COMSOL Multiphysics to obtain the strain from a realistic gate-stack of planar hole quantum dot confinement. Recently, spin-orbit coupling terms $\propto k$ have been shown to be induced by strain inhomogeneity.\cite{abadillo2023hole} Our hybrid approach to realistic device modeling suggests that strain inhomogeneity due to both random alloy disorder and gate-induced strain make a strong contribution to the linear-$k$ Dresselhaus spin-orbit coupling, which eventually dominates hole spin EDSR; and there exist specific in-plane orientations of the global magnetic field $\mathbf{B}$ and the microwave drive $\Tilde{\mathbf{E}}_{\text{ac}}$ for maximum EDSR Rabi frequency of the hole spin qubit. The current model including strain inhomogeneity accurately predicts the EDSR Rabi frequency to be $\!\sim\!100$ MHz for typical electric and magnetic fields in experiments, which represents at least an order of magnitude improvement in accuracy over phenomenological models assuming uniform uniaxial strain. State-of-the-art atomistic tight binding calculations via nano-electronic modeling (NEMO3D) are in agreement with the $\mathbf{k}{\cdot}\mathbf{p}$ description. 
\end{abstract}



\section{Introduction}\label{sec:introrough}
 Quantum computing\cite{benioff1980computer, feynmann1982simulating, deutsch1985quantum} with semiconductor spin qubits\cite{kane1998silicon, loss1998quantum, Elzerman2004} has burgeoned over the past two decades,\cite{petta2005coherent,Hanson2007,chatterjee2021semiconductor} owing to their prospective scalability from the elementary few-qubit systems to large arrays.\cite{Vandersypen2017, lawrie2020quantum, van2021two} Quantum dots\cite{Divincenzo2000} synthesized in Group IV materials are good candidates for spin qubits, \cite{zwanenburg2013silicon,scappucci2021germanium} having the advantage of excellent compatibility with the semiconductor fabrication industry.\cite{Vandersypen2017} Silicon and germanium are of particular interest, as the possibility of having net-zero nuclear spin via isotopic purification,\cite{itoh1993ge,itoh2003si,yoneda2018quantum,tyryshkin2012electron,prechtel2016decoupling,bosco2021fully} and the absence of piezoelectric phonons,\cite{cardona2005fundamentals,relaxationTahan2014} can lead to long qubit coherence. 
 Recently, quantum computing with holes in Si and Ge has garnered tremendous interest,\cite{hu2012hole,chesi2014controlling,maurand2016cmos,bohuslavskyi2016pauli,vuku2018single,hendrickx2018gate,lawrie2020spin,liles2021electrical,piot2022single,Camenzind2022_4K} dovetailing with the collective effort towards achieving all-electrical control of spin qubits.\cite{pioro2008electrically,pribiag2013electrical,voisin2016electrical,srinivasan2016electrical,hung2017spin,marcellina2018electrical,venitucci2019simple,gao2020site,liles2021electrical} Compared to magnetic fields, electric fields are easier to apply locally, and enable low-power qubit control with faster gate speeds. Holes in the $p$-type valence band (VB) of Si and Ge are characterized by an effective spin-$3/2$ and exhibit a large intrinsic spin-orbit coupling (SOC).\cite{rashba1988spin,winkler2003spin, winkler2004spin, Culcer_Precession_2006, marcellina2017spin, Katsaros2020ZeroField} The role of SOC in all electrical qubit control is well established.\cite{golovach2006electric, nadj2010spin, tanttu2019controlling, adelsberger2022hole} Germanium as a material platform exhibits strong Fermi-level pinning close to the valence band edge,\cite{dimoulas2006fermi,sammak2019shallow} which means that hole qubits can be realized within reasonable gate bias ranges in experiments. Additionally, this enables the formation of Ohmic contacts for semiconductor-superconductor hybrid applications,\cite{li2018coupling,hendrickx2018gate,vigneau2019germanium,xu2020dipole,Aggarwal2021EnhanceSC,yu2022strong,valentini2023radio} playing a significant factor in scaling up spin qubit systems.
 

Spin-orbit coupling in hole systems mediates anisotropy and tunability of the $g$-tensor,\cite{voisin2016electrical, liles2021electrical, froning2021strong} which can have beneficial effects for qubit coherence.\cite{wang2021optimal, bosco2021sweetspots, piot2022single, hendrickx2024sweet} Therefore for fast and coherent spin qubit manipulation holes may offer advantages\cite{scappucci2021germanium, fang2022recent} over electron systems, where operational speed is often sacrificed for coherence \cite{Awschalom2007, yang2019silicon}, and reliance on micromagnets complicates scaling up \cite{pioro2008electrically, kawakami2014electrical, zajac2018resonantly, klemt2023electrical}.
Strained germanium quantum wells (QW)\cite{mizokuchi2018ballistic, stehouwer2023germanium,dsouza2024heavy} and silicon complementary metal semiconductor oxide (CMOS)\cite{bohuslavskyi2016pauli, wei2020estimation} are particularly suited for high frequency hole qubit experiments demonstrating fast and coherent electrical control.\cite{ares2013nature, pribiag2013electrical, maurand2016cmos, hendrickx2018gate, wang2022ultrafast, jirovec2021singlet, Geyer2022Exchange, piot2022single}
 Strained Ge QWs in undoped heterostructures have been at the forefront of the recent experimental effort in holes,\cite{sammak2019shallow, vuku2018single} ranging from universal single qubit and two-qubit logic,\cite{lawrie2020spin, hendrickx2020single, Hendrickx2020} singlet-triplet hole qubit,\cite{jirovec2021singlet} a four-qubit germanium processor\cite{hendrickx2021four} to shared control of large crossbar QD array.\cite{Borsoi2022-16QDs} With high hole mobilities of ${\approx}10^6\,\text{cm}^2/(\text{V}{\cdot}\text{s})$\cite{dobbie2012ultra,lodari2022lightly} and a low effective mass of $0.05\,m_e$,\cite{lodari2019light} strained Ge quantum wells in $\text{Ge}/\text{Si}_{1-x}\text{Ge}_{x}$ heterostructures allow the formation of large, low-disorder hole quantum dots.\cite{lodari2021low}$\,\,$
 
Whereas the theory of Ge and Si hole qubits has evolved apace, with advances in the understanding of electrically driven spin resonance (EDSR),\cite{bulaev2007electric, kloeffel2018direct, bosco2023phase} coherence,\cite{fischer2008spin, wang2021optimal, mauro2024geometry} relaxation,\cite{hu2012hole, Shalak2023HoleRTN} $g$-factor anisotropy and singlet-triplet qubit control\cite{hung2017spin,Qvist2022Aniso-g}, the variations in dot size, strain distribution, and energy dispersion across the many existing qubit architectures have made it difficult to develop a unified theoretical approach. Experiments have shown a high variability in the properties of hole spin qubits, which is believed to be due to the inhomogeneous strain field to which they are exposed, as revealed by X-ray diffraction microscopy.\cite{corley2023nanoscale} Inhomogeneous strain in Ge quantum dots, which arises from random alloy disorder, lattice mismatch at the interface and gate electrode contraction, leads to additional inversion asymmetry.\cite{sherman2005spin,bindel2016probing,secchi2024envelope} The resultant spin-orbit coupling is extremely important for realistic device modeling, as it can overwhelm the bare spin-orbit coupling from the Luttinger Hamiltonian.\cite{abadillo2023hole,rodriguez2023linear,secchi2024envelope}. It results, for example, in $g$-factor anisotropy and $g$-tensor modulation, qubit energy level variations,\cite{terrazos2021theory} and anisotropic noise sensitivity.\cite{hendrickx2024sweet} In the context of silicon holes, Ref.~\citenum{liles2021electrical} showed that non-uniform strain arising from thermal contraction of the gate electrodes has a sizable effect on the SOC and g-tensor in hole quantum dots. At the same time recent analytical and numerical models have captured the non-trivial interplay between intrinsic spin-orbit coupling and external electrical control.\cite{adelsberger2022hole, sarkar2023electrical,wang2024electrical} However, qubit operational timescales remain underestimated in theory compared to experimental observations. Moreover, a flexible approach that can systematically accounts for the device-specific local non-uniformity is at present lacking in the field. For example, tight binding simulations can reasonably describe the atomistic strain inhomogeneity in Ge/Si$_{1-x}$Ge$_x$ heterostructure, but the key contribution to the spin-orbit coupling of the gate electrodes is overlooked. At the same time, while the linear-$k$ SOC in Ge hole QDs resulting from strain gradients has recently been investigated in Ref.~\citenum{rodriguez2023linear} using the finite element method (FEM), this description is also incomplete without an accurate account of atomic-scale inhomogeneities.

 \begin{figure}[tbp]
\subfloat[]{\begin{minipage}[c]{0.56\linewidth}
\vspace{-0.06in}
\includegraphics[width=\linewidth]{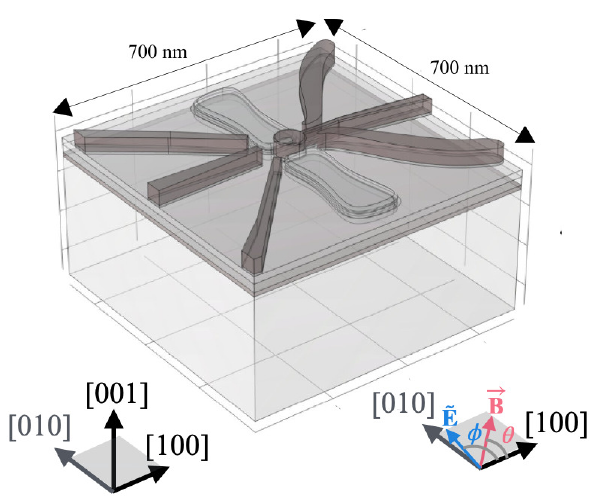}
\label{fig:geomGateStackFull}
\end{minipage}}
\hfill
\subfloat[]{\begin{minipage}[c]{0.44\linewidth}
\includegraphics[width=\linewidth]{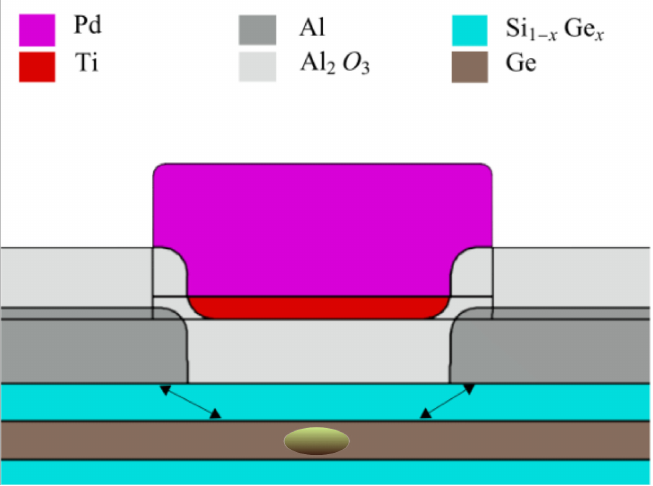}
\label{fig:GeomSideView}
\end{minipage}}
\vspace{-0.1in}
\caption{{\bf Design for a Ge hole quantum dot} a) with realistic gatestack, b) cross-section of the design showing the 10 nm thick Ge layer between the lower and upper Ge/Si$_{0.2}$Ge$_{0.8}$ layers of thickness 60 nm and 10 nm, respectively. The first layer of Al source and drain gates (20 nm thick) sit atop the heterostructure, followed by a dielectric Al$_2$O$_3$ layer (17 nm), and a second layer of Pd/Ti electrodes (5/35 nm). The hole quantum dot forms directly under the Pd/Ti top gate(TG), as indicated by the oval shape in the Ge layer in b). The angular orientations of the applied in-plane magnetic field $\mathbf{B}$ and the ac electric field $\tilde{E}_{\text{ac}}$ with respect to the $[100]$ crystallographic axis is denoted by $\theta$ and $\phi$, respectively.}
\label{fig1:sketch}
\vspace{-0.1 in}
\end{figure}

In light of these challenges, in this work we adopt a device-specific approach to describe comprehensively the effect of realistic strain inhomogeneity on coherent control of planar Ge hole qubits. In our approach, the strain originating from the random alloy disorder is obtained using the valence force field method, which calculates the equilibrium position of the atoms in a multimillion atom Ge/Si$_{1-x}$Ge$_x$ heterostructure. The strain due to the gate electrodes atop the heterostructure is obtained from the thermal expansion modeling of the device in COMSOL Multiphysics. The resultant atomistic strain nonuniformity is added to the $\mathbf{k}\,{\cdot}\,\mathbf{p}$ effective mass model of the hole quantum dot. Our methodology can incorporate both perpendicular (out-of-plane) and parallel (in-plane) magnetic field operations of the hole-spin qubit, however, the main focus of this work will be on the in-plane magnetic field operation of Ge hole qubits. This choice is motivated by the recent experiments in Refs.~\citenum{wang2024operating, hendrickx2024sweet}, where it is established that the in-plane alignment of the applied $\mathbf{B}$ field plays a significant role in optimizing the coherence of Ge hole spin qubits. Note that, in contrast to the large bare Zeeman splitting in an out-of-plane applied $\mathbf{B}$, the small $g$-factor of holes in an in-plane magnetic field entirely originates from spin-orbit coupling,\cite{sarkar2023electrical, wang2024electrical} which in turn renders the $g$-tensor engineering of holes extremely sensitive to the precision of the $\mathbf{B}$ field orientation in experiments.\cite{van2024coherent,wang2024operating}

The central findings of this work are as follows: i) while the semi-analytical model incur much less computational cost than a full numerical tight binding or FEM analysis, it enables the addition of both short-range atomistic and long-range gate induced strain for the calculated single hole qubit Rabi frequency to benchmark well against experiments, ii) the calculated qubit Rabi frequency shows important variation with the respective orientation between the applied in-plane magnetic field $B_\parallel$ and the alternating ac electric field $\hat{E}_{\text{ac}}$, mediated by SOC due to both random alloy disorder and gate-induced strain.

The outline of this paper is as follows. In the section titled '{\it $\mathbf{k}\cdot\mathbf{p}$ model with inhomogeneous strain }', the effective mass $\mathbf{k}\cdot\mathbf{p}$ model is described, focusing on the implications of the uniaxial strain assumption on planar Ge hole qubit EDSR. Next, the tight binding simulation of the planar Ge hole quantum dot atomistic wave function is outlined, and the resultant EDSR calculation is presented. A full strain profile is developed subsequently. In the section titled '{\it Hole qubit operation in the presence of inhomogeneous strain}', the key mechanism of full-electrical qubit control in the presence of nonuniform strain is discussed. We end with our conclusions and outlook.
\begin{figure*}[htbp!]
\subfloat[]{\begin{minipage}[c]{0.25\linewidth}
\includegraphics[width=0.95\linewidth]{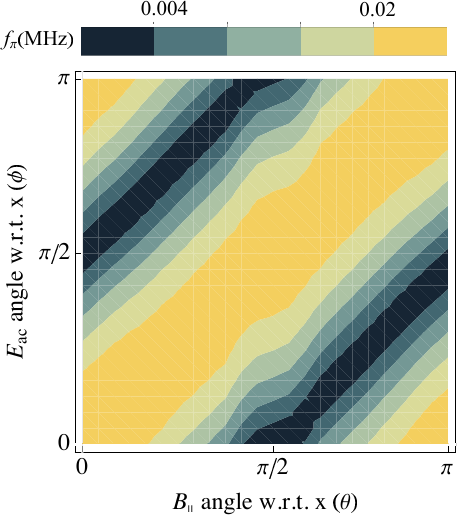}
\label{fig:EDSR20nmnorough200mT1p5}
\end{minipage}}
\hfill
\subfloat[]{\begin{minipage}[c]{0.25\linewidth}
\includegraphics[width=0.95\linewidth]{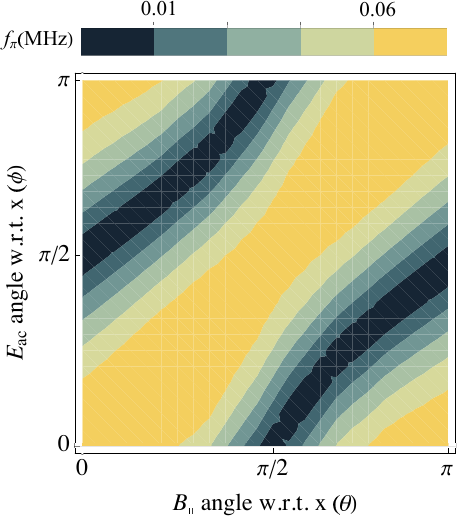}
\label{fig:EDSR20nmnorough670mT1p5}
\end{minipage}}
\hfill
\subfloat[]{\begin{minipage}[c]{0.25\linewidth}
\includegraphics[width=0.95\linewidth]{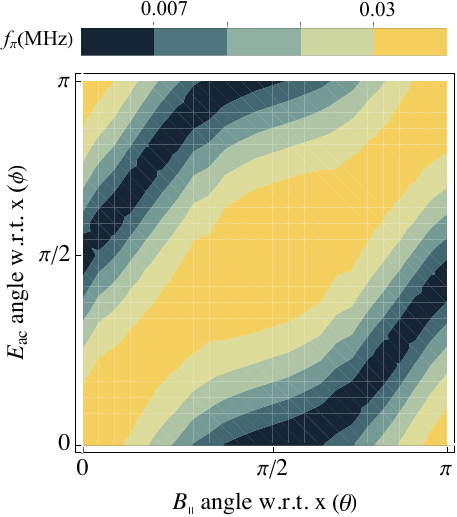}
\label{fig:EDSR45nmnorough200mT1p5}
\end{minipage}}
\hfill
\subfloat[]{\begin{minipage}[c]{0.25\linewidth}
\includegraphics[width=0.95\linewidth]{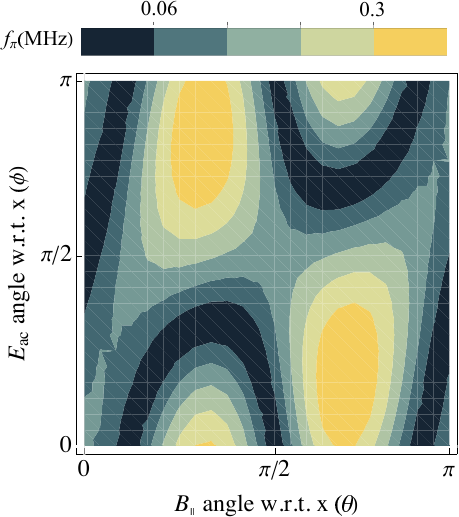}
\label{fig:EDSR45nmnorough670mT1p5}
\end{minipage}}
\hfill
\vspace{-0.1in}
\subfloat[]{\begin{minipage}[c]{0.25\linewidth}
\includegraphics[width=0.95\linewidth]{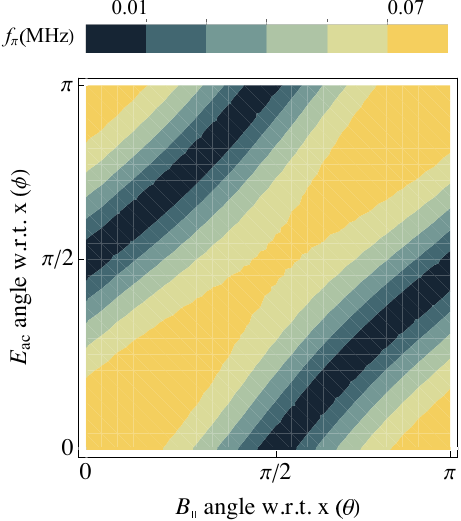}
\label{fig:EDSR20nmnorough200mT}
\end{minipage}}
\hfill
\subfloat[]{\begin{minipage}[c]{0.25\linewidth}
\includegraphics[width=0.95\linewidth]{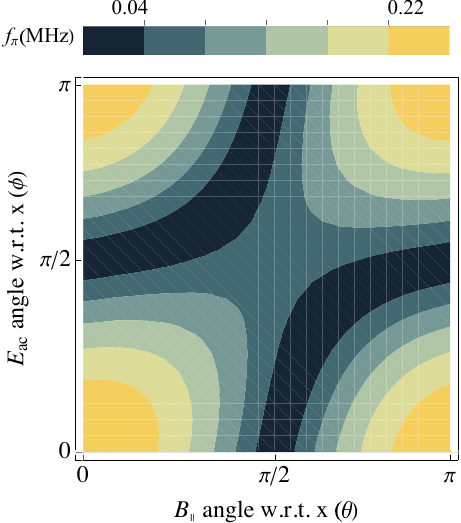}
\label{fig:EDSR20nmnorough670mT}
\end{minipage}}
\hfill
\subfloat[]{\begin{minipage}[c]{0.25\linewidth}
\includegraphics[width=0.95\linewidth]{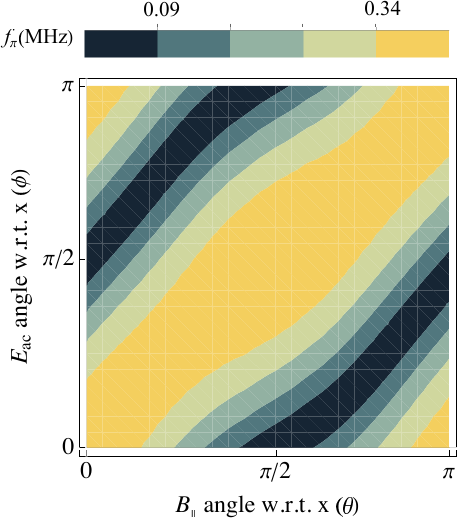}
\label{fig:EDSR45nmnorough200mT}
\end{minipage}}
\hfill
\subfloat[]{\begin{minipage}[c]{0.25\linewidth}
\includegraphics[width=0.95\linewidth]{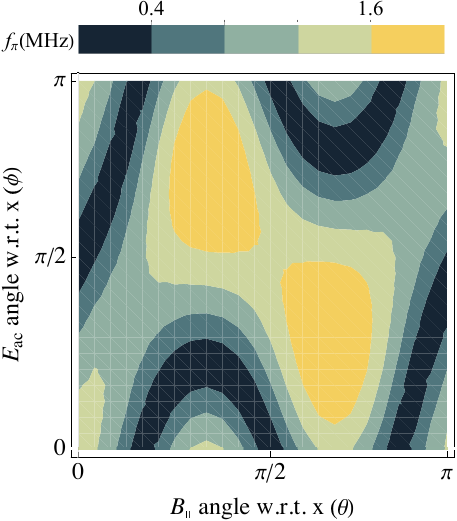}
\label{fig:EDSR45nmnorough670mT}
\end{minipage}}
\caption{{\bf Hole qubit EDSR Rabi frequency $f_\pi$ calculated from the k.p model under uniform uniaxial strain.} The angles of $\mathbf{B}$ and $\Tilde{\mathbf{E}}_{\text{ac}}$ w.r.t. the [100] axis are denoted by $\theta$ and $\phi$, respectively. At $F_z{=}1.5$ MV/m (a-d) and  $F_z{=}15$ MV/m (e-h)top-gate fields, with uniaxial strain assumption, $f_\pi$ is evaluated for a smaller dot first, a),e) $a_d{=}20$ nm, $\lvert B\rvert{=}200$ mT, b),f) $a_d{=}20$ nm, $\lvert B\rvert{=}670$ mT; and then for a larger dot, c),g) $a_d{=}45$ nm, $\lvert B\rvert{=}200$ mT, d),h) $a_d{=}45$ nm, $\lvert B\rvert{=}670$ mT. Here $a_d$ signifies the dot radius, and $\lvert B\rvert$ denotes the magnitude of the applied magnetic field.  As suggested by the color bars, the EDSR Rabi frequency improves with increasing the dot radius as well as the applied $\mathbf{B}$ strength. $f_\pi(\theta,\phi)$ largely follows the trend of the dot product $\mathbf{B}\cdot\Tilde{\mathbf{E}}_{\text{ac}}$. As shown in d),h), for large dot at higher magnetic field the strong orbital vector potential terms cause the $f_\pi(\theta,\phi)$ trend to deviate from the dot product $\mathbf{B}\cdot\Tilde{\mathbf{E}}_{\text{ac}}$ relation. The microwave drive amplitude is given by $E_{\text{ac}}{=}10$ kV/m.}
\label{fig2:Uniaxial}
\vspace{-0.3 cm}
\end{figure*}
\section{Results and Discussion}
\subsection{$\mathbf{k}\cdot\mathbf{p}$ model with inhomogeneous strain}
\subsubsection{Luttinger-Kohn Hamiltonian}\label{sec:3dkdotp}

The $\mathbf{k}\cdot\mathbf{p}$ formalism describes the energy dispersion of holes in the topmost valence bands of III-V semiconductors. The resultant Luttinger-Kohn Hamiltonian\cite{luttinger1955} has the form: 
\begin{eqnarray}\label{eq:1HLK}
    H_{LK}(k)\!&{=}\!&\frac{\hbar^2}{2 m_0}\left[\gamma_1 k^2-2\gamma_2 \left((J_x^2-(1/3)J^2)k_x^2+c.p.\right)\right.\nonumber\\
   &&\left.-4\gamma_3\left(\left\{k_x,k_y\right\}\left\{J_x,J_y\right\}+c.p.\right)\right]
\end{eqnarray}
where $J_x,\,J_y,\,J_z$ signify the spin-3/2 Dirac matrices, $\gamma_1,\,\gamma_2,\,\gamma_3$ are the Luttinger parameters, 'c.p.' stands for cyclic permutation of the indices $x{\rightarrow}y{\rightarrow}z{\rightarrow}x$, and anticommutation is denoted by $\{A,B\}{=}(AB+BA)/2$. The six-fold degeneracy of the VB at the $\Gamma$ point is lifted due to the intrinsic spin-orbit coupling ($l{=}1$, $s{=}1/2$). The resultant total angular momentum eigenstates: four-fold $j{=}3/2$ and two-fold $j{=}1/2$ are separated by the large spin-orbit gap of Ge ($\Delta_0{=}325$ meV). The topmost VB states in the $\left|j,m_j\right\rangle$ basis $\left\{\left|\frac{3}{2} \frac{3}{2}\right\rangle_{\text{HH}},\left|\frac{3}{2} -\frac{3}{2}\right\rangle_{\text{HH}},\left|\frac{3}{2} \frac{1}{2}\right\rangle_{\text{LH}},\left|\frac{3}{2} -\frac{1}{2}\right\rangle_{\text{LH}}\right\}$ are given by the $4{\times}4$ picture in Eqn.~\ref{eq:1HLK}. The subscripts signify the heavy-hole (HH) and light-hole (LH) sub-bands at finite $\mathbf{k}$. In presence of an external in-plane magnetic field, the single hole quantum dot spin qubit Hamiltonian is:
\begin{equation}\label{eq:2QDH}
H_{QD}=H_{LK}\!\!\left(k+eA/\hbar\right)+V_{\text{conf}}+eF_zz+H_Z+ H_\varepsilon
\end{equation}
where $V_{\text{conf}}$ is the Ge quantum dot (QD) confinement due to the gates and the type-I band alignment in the Ge/Si$_{1-x}$Ge$_x$ heterostructure. The QD is defined below the top gate (TG), the electric field of which is denoted by $F_z$. The external $\mathbf{B}$ induced $H_Z{=}2\kappa\mu_B\mathbf{B}{\cdot}\mathbf{J}{+}2q\mu_B\mathbf{B}{\cdot}\mathcal{J}$ consists of the bulk isotropic ($\kappa$) and anisotropic ($q$) Zeeman interactions. Note that $\mathbf{J}{=}(J_x,J_y,J_z)$ and $\mathcal{J}{=}(J_x^3,J_y^3,J_z^3)$. The vector potential $\mathbf{A}$ due to the applied $\mathbf{B}$ rectifies the canonical momentum $\mathbf{k}$, giving rise to additional spin-orbit coupling terms.

The strain Hamiltonian $H_\varepsilon$ originates due to the transformation between strained $r'$ and unstrained $r$ coordinates:$\,\,r'_\alpha{=}r_\alpha{+}\sum_\beta\varepsilon_{\alpha\beta}r_\beta$. This is required to describe the strained Ge band structure in terms of the known $\mathbf{k}{\cdot}\mathbf{p}$ material parameters of unstrained Ge. One can calculate $H_\varepsilon$ in the lowest order of perturbation theory as the Bir-Pikus-Luttinger-Kohn Hamiltonian:
\begin{eqnarray}\label{eq:3strainHam}
    H_\varepsilon&=&a\,{\it tr}\,\varepsilon+b\left((J_x^2-(1/3)J^2)\varepsilon_{xx}+c.p.\right)\nonumber\\
    &&+d/\sqrt{3}\left(2\left\{J_x,J_y\right\}\varepsilon_{xy}+c.p.\right)
\end{eqnarray} 
where ${\it tr}\,\varepsilon$ signifies the trace of the strain tensor with components $\varepsilon_{\alpha\beta}$. Note that in the case of spatially varying strain, there exist additional diagonal terms in Eqn.~\ref{eq:3strainHam}, proportional to $\nabla^2\varepsilon_{\alpha\alpha}(\mathbf{r})I_{4\times4}$, $\{k_\alpha,\varepsilon_{\alpha\beta}(\mathbf{r})\}k_\beta I_{4\times4}$ and $\varepsilon_{\alpha\beta}(\mathbf{r}) k_\alpha k_\beta I_{4\times4}$, respectively.\cite{secchi2024envelope} We have explicitly checked that these terms make a negligible contribution to our results. Moreover, our results derived based on the $\mathbf{k}\cdot\mathbf{p}$ model incorporating Eqn.~\ref{eq:3strainHam} are in good agreement with the tight-binding calculation; which, in contrast to the Bir-Pikus picture, includes the full inhomogeneity of the strain profile. The deformation potentials have the following values: $a{=}-2\,\text{eV}$, $b{=}-2.16\,\text{eV}$, $d{=}-6.06\,\text{eV}$.\cite{wortman1965young}

In a planar Ge hole quantum dot such as the gate-defined QD in Fig.~\ref{fig:geomGateStackFull}, the out-of-plane QD confinement is much stronger than the lateral confinement. This produces a large heavy hole-light hole gap $\Delta_{HL}{\sim\,50}$ meV, and the QD ground state becomes predominantly heavy-hole in nature. The structural inversion asymmetry (SIA) originates from the top gate field $F_z$, which in conjunction with the intrinsic SOC and the external $\mathbf{B}$ induced Zeeman and orbital terms, contributes to the fast electrical control of the hole spin qubit.\cite{sarkar2023electrical} 
\begin{figure*}[htbp!]
\subfloat[]{\begin{minipage}[c]{0.48\linewidth}
\vspace{-0.07in}
\includegraphics[width=\linewidth,height=3.2in]{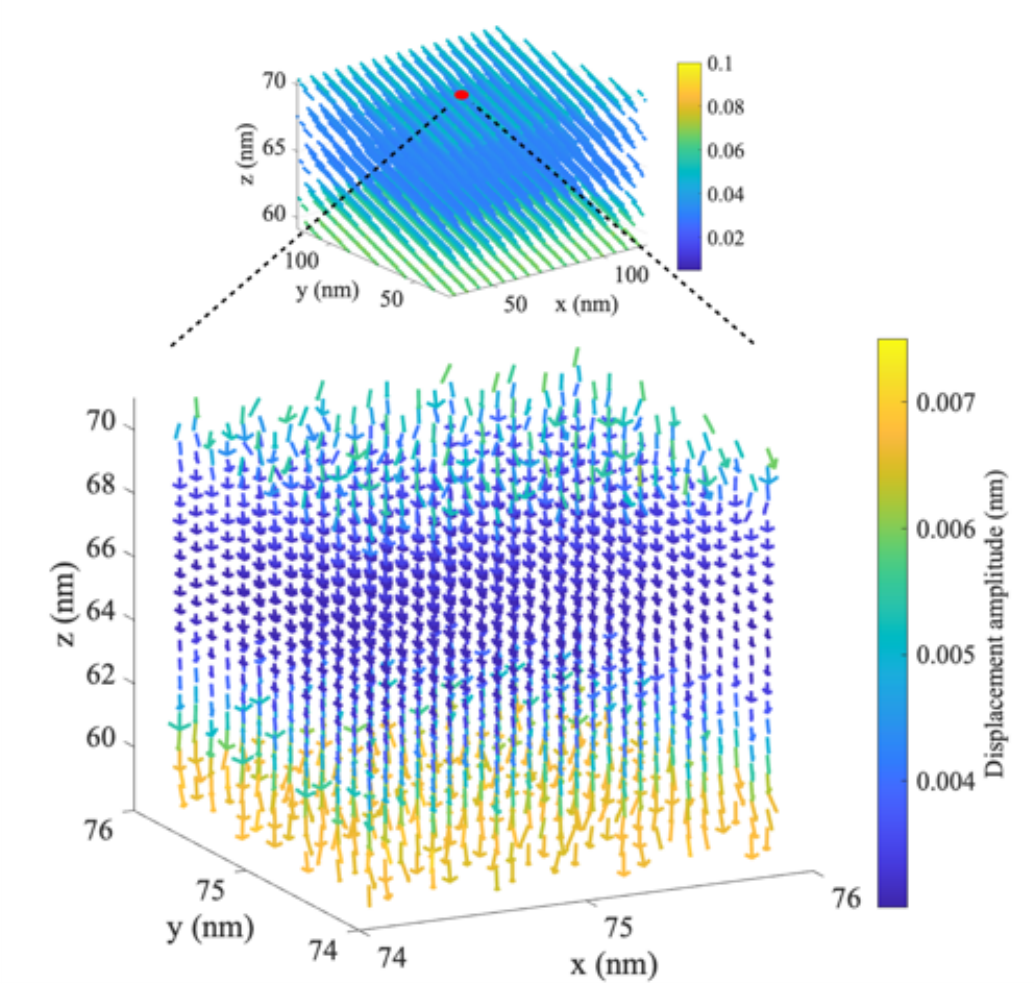}
\label{fig:displacementprofile}
\end{minipage}}
\hfill
\subfloat[]{\begin{minipage}[c]{0.27\linewidth}
\vspace{0.2in}
\includegraphics[width=\linewidth]{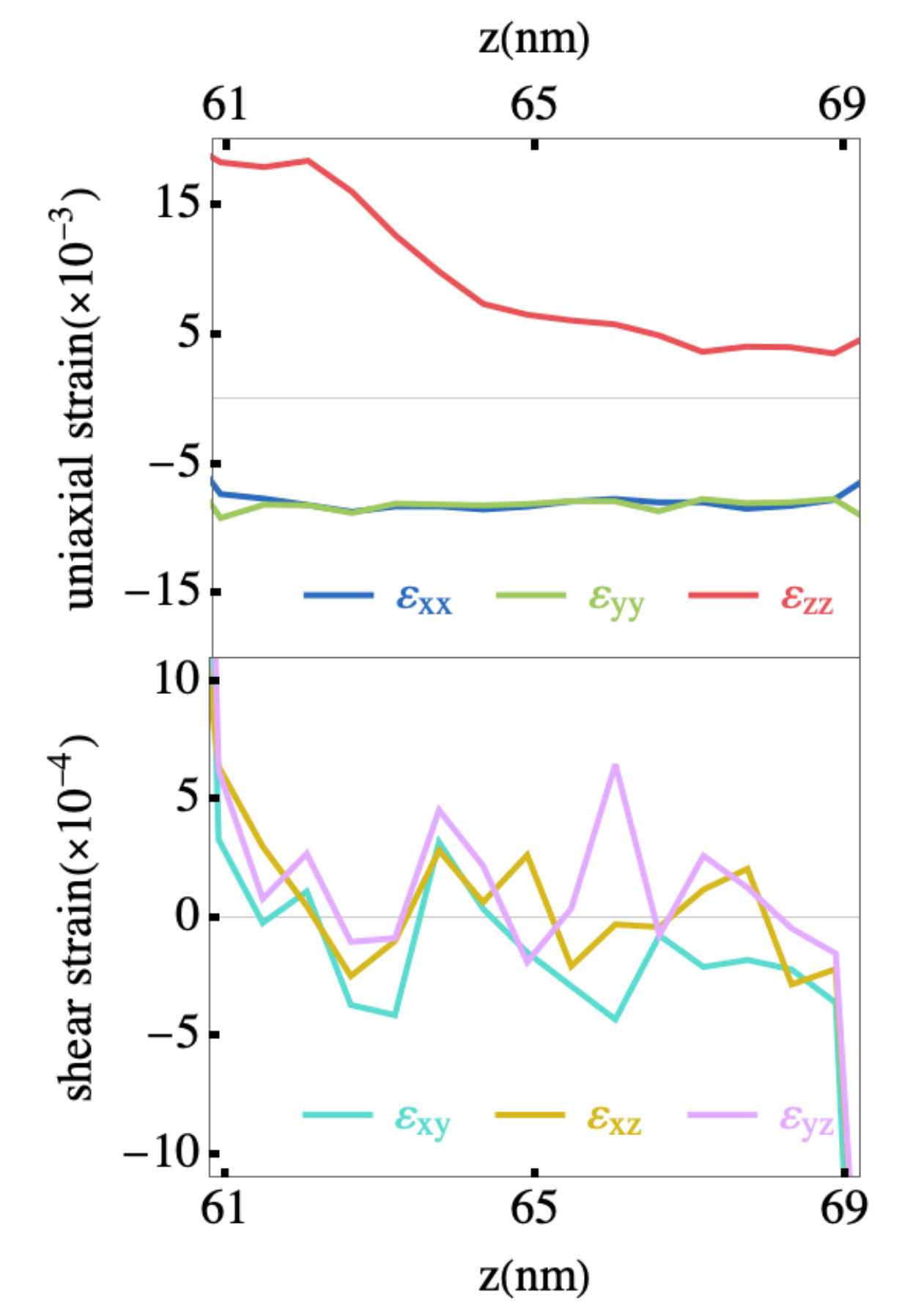}
\label{fig:strainZcuts2D}
\end{minipage}}
\hfill
\subfloat[]{\begin{minipage}[c]{0.25\linewidth}
\vspace{-0.07in}
\includegraphics[width=0.9\linewidth]{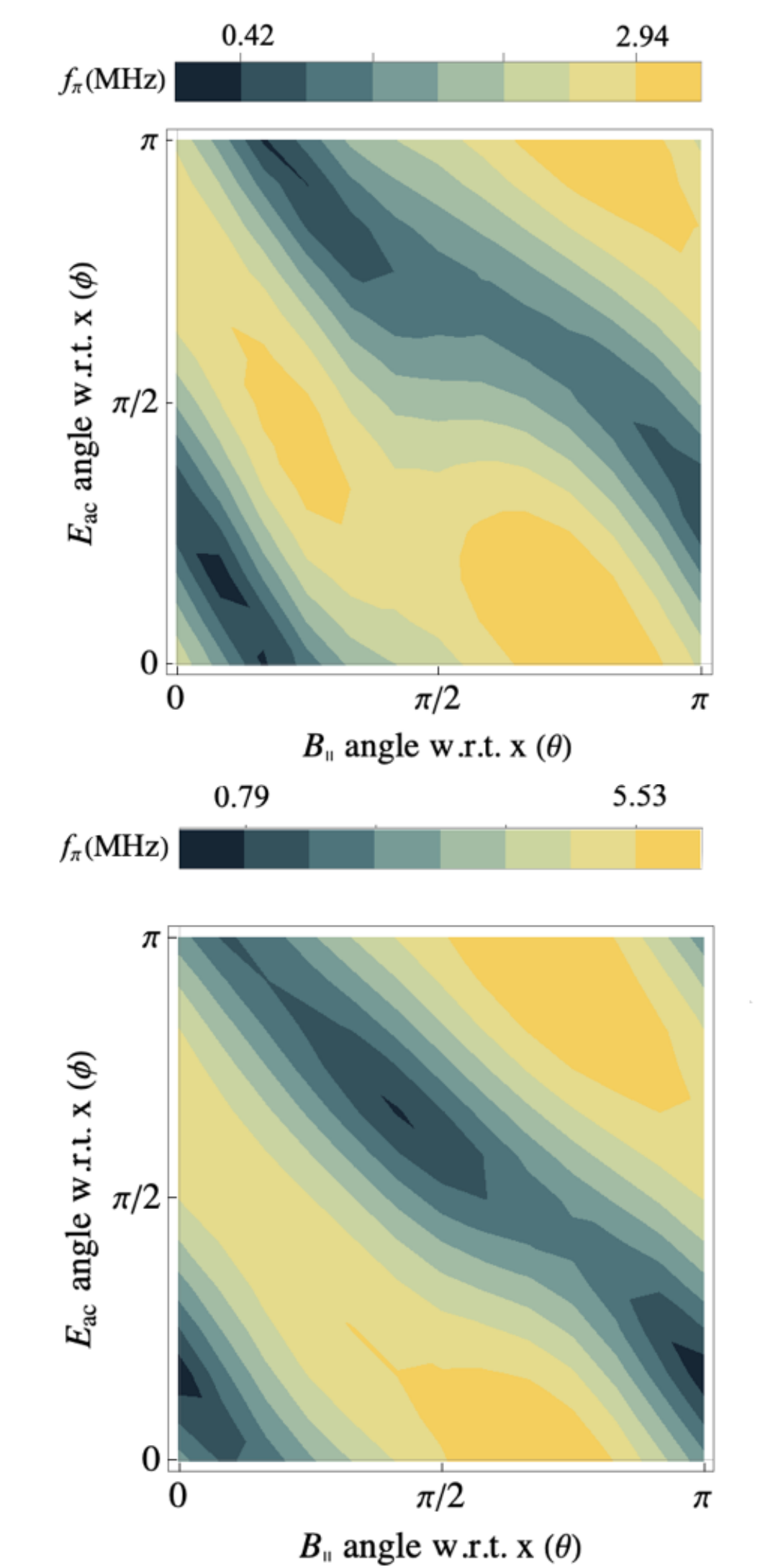}
\label{fig:fEDSRBdotEfuncplotNEMO}
\end{minipage}}

\caption{{\bf tight binding simulation of the heterostructure profile in NEMO3D} a) with random alloy disorder induced displacement vector field data of Ge atoms, calculated via the valence force field method. On the top panel, displacement amplitudes over a $150\times150\times10$ $\text{nm}^3$ box around the quantum dot with in-plane radius $a_d{=}45$ nm and quantum well width $L_z{=}10$ nm is shown. A small section at the center of the dot is magnified to elaborate the randomness close to the buffer interfaces, as well as the variation of the displacement magnitudes across the quantum well. b) Variation of the uni-axial strain tensor components $\varepsilon_{xx},\,\varepsilon_{yy},\,\varepsilon_{zz}$ and shear strain tensor components $\varepsilon_{xy},\,\varepsilon_{xz},\,\varepsilon_{yz}$ with $z$ (across the QW). c) NEMO3D simulation of hole EDSR Rabi frequency at the top-gate field $F_z{=}1.5$ MV/m (top) and $F_z{=}15$ MV/m (bottom). The external magnetic field is $\lvert B\rvert{=}670$ mT. The $f_\pi(\theta,\phi)$ trend shows stark difference compared to the uniaxial case. The microwave drive amplitude is $|E_{\text{ac}}|{=}10$ kV/m.}
\label{fig3:NEMO}
\vspace{-0.3 cm}
\end{figure*}

To evaluate the wave function of the hole quantum dot spin qubit, $H_{QD}$ is expanded in the $\{\psi_{l}(x),\,\psi_{m}(y),\,\psi_n(z)\}$ basis; which comprises of the eigenstates of parabolic potential of radius $a_d$ in x-y, and infinite well potential in z of length $L_z{=}10$ nm. The $n,\,m,\,l$ indices take integer values. Convergence of results for different gauge choices is achieved for a simulation size of $n,\,l,\,m\,\in[1,4]$.\cite{sarkar2023electrical} The final $256{\times}256$ matrix is numerically diagonalized to calculate the energy levels of the hole quantum dot: $H_{QD}\left|\Psi_{QD}\right\rangle{=}\lambda_E\left|\Psi_{QD}\right\rangle$. The heavy hole type qubit levels can be expressed as $\left|\Psi_{QD}^{GS}\right\rangle=\left(f(x,y,z)\left|\frac{3}{2}\right\rangle+\text{LH admixtures}\right)$ and $\left|\Psi_{QD}^{ES}\right\rangle=\left(g(x,y,z)\left|-\frac{3}{2}\right\rangle+\text{LH admixtures}\right)$, with Zeeman splitting $\Delta\mathbf{E}$. The atomistic hole qubit wave function can be evaluated via a tight binding calculation, as outlined in the next section.

An applied a.c. electric field $e\Tilde{\mathbf{E}}_{\text{ac}}(t){\cdot}\mathbf{r}$ in resonance with the Larmor frequency $\nu{=}\Delta\mathbf{E}/h$, $\pi$-rotates the hole spin via the electric dipole spin resonance (EDSR). The EDSR Rabi frequency is calculated as $f_\pi=\left\langle\Psi_{QD}^{GS}\right|e\Tilde{\mathbf{E}}_{\text{ac}}(t){\cdot}\mathbf{r}\left|\Psi_{QD}^{ES}\right\rangle$. Following the Fig.~\ref{fig:geomGateStackFull} notations for $\mathbf{B}{=}B_\parallel\left( \cos\theta,\,\sin\theta,\, 0 \right)$ and $\Tilde{\mathbf{E}}_{\text{ac}}(t){=}E_{ac}\cos (2\pi f_L t)\left(\cos\phi,\,\sin\phi,\, 0\right)$, and considering the simple case without inhomogeneity, it can be assumed that the strain in the 10-nm-wide quantum well is uniaxial with $\varepsilon_{xx}{=}\varepsilon_{yy}$=-0.6\%. The calculated EDSR Rabi frequency for a smaller hole quantum dot of radius $a_d{=}20$ nm is maximum when $\mathbf{B}\parallel\Hat{\mathbf{E}}_{\text{ac}}$, for both low ($\lvert B\rvert$=200 mT) and high ($\lvert B\rvert$=670 mT) magnetic field strength (Figs.~\ref{fig:EDSR20nmnorough200mT}, \ref{fig:EDSR20nmnorough670mT}). In other words, the angular orientation of the in-plane magnetic field and the ac microwave drive w.r.t. $x\parallel[100]$, given by $\theta$ and $\phi$, respectively, affects the EDSR Rabi frequency through the scaler product relation: $f_\pi\propto\cos(\theta-\phi)$.

The EDSR Rabi frequency can be calculated as $f_\pi{=}\frac{\alpha_{R3}\mathbf{B}\cdot\Tilde{\mathbf{E}}_{\text{ac}}}{\Delta^2}$ in the lowest energy heavy-hole subspace (hh1) using the Schrieffer-Wolff transformation,\cite{winkler2003spin,kloeffel2018direct,sarkar2023electrical} where the spin-orbit coefficient $\alpha_{R3}$ originates from the SIA $k$-cubic Rashba 2D Hamiltonian: $\alpha_{R3}\left(\{k_+^2,k_-\}\sigma_+{-}\{k_-^2,k_+\}\sigma_-\right)$; and $\Delta$ denotes the splitting between the qubit levels (hh1) and the next orbital levels (hh2). For a larger dot of radius $a_d{=}45$ nm, the EDSR Rabi frequency complies with the dot product formula at low $\lvert B\rvert$ (Fig.~\ref{fig:EDSR45nmnorough200mT}), however, $f_\pi(\theta,\phi)$ is non-trivial at higher $\lvert B\rvert$ (Fig.~\ref{fig:EDSR45nmnorough670mT}). This is due to the vector potential terms, which scale linearly with the magnetic field stregth and the QD size (e.g. $\mathbf{A}{=}\frac{1}{2}\mathbf{B}\times\mathbf{r}$ in the symmetric gauge), and have a sizable effect on the spin-orbit coupling mediated hole qubit operation of large Ge hole QDs at high magnetic field. As described in ref.~\citenum{sarkar2023electrical}, the quasi-2D limit would be unable to fully capture the orbital-B effect in the lowest order of perturbation theory, due to the interplay of out-of-plane and in-plane degrees of freedom.
\begin{figure*}[htbp!]
\vspace{-0.3 cm}
    \centering
    \includegraphics[scale=0.8,trim= 0.2 45 0.2 45,clip]{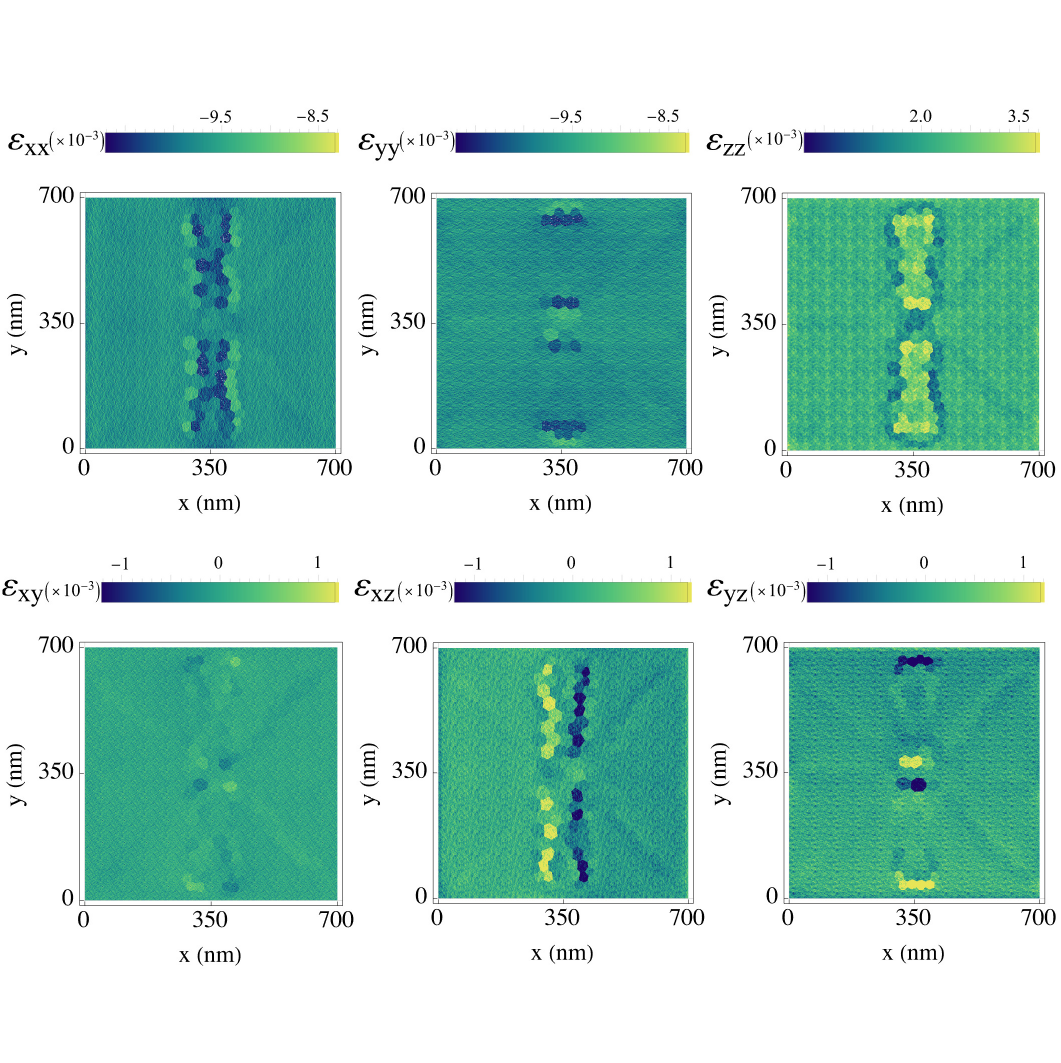}
    \caption{{\bf Cumulative strain profile in realistic Ge hole quantum dot.} The spatial profile of the strain tensor components $\varepsilon_{ij}(x,y)$ in the Ge layer (z=60 nm cut) of the device in Fig.~\ref{fig1:sketch} due to random alloy disorder as well as gate contraction. The features along the $y$-axis here are due to the Al source and drain electrodes as shown in the Fig.~\ref{fig1:sketch} schematic. The atomic granularity comes from the random alloy disorder; while the long-range feature due to the gate contractions is signified by the COMSOL mesh data superimposed atop the atomistic VFF data.}
    \label{fig4:strainmaps}
    \vspace{-0.3 cm}
\end{figure*}

\subsubsection{Tight-binding calculation}\label{sec:TB}

The assumption of uniaxial strain in the previous section has limited applicability in realistic devices, where hole qubit operation is significantly influenced by local features in the vicinity of the quantum dot. Using the tight binding (TB) method, the hole qubit wave function can be calculated as a linear combination of atomic orbitals at each lattice site. Here, the multimillion atom 3D nano-electronic modeling tool (NEMO3D) is implemented to calculate the hole QD eigen-energies and the eigen-functions, which relies on a nearest-neighbor empirical tight binding model.\cite{klimeck2007atomistic} The valence force field calculation is incorporated into the NEMO3D tool, such that the resultant wave function encapsulates the effect of random alloy disorder.

The tight binding methodology entails selecting a basis consisting of different orbitals (such as s,p,d and $s^*$ in case of $sp^3d^5s^*$) centered around each atom, which also form the basis for the Hamiltonian which assumes the form:
\begin{equation}
\hat{H}=\sum_{i}\epsilon_i^{\nu}c_{i,{\nu}}^{\dagger}c_{i,{\nu}}+\sum_{i,\nu\mu}t_i^{\nu,\mu}c_{i,{\nu}}^{\dagger}c_{i,{\mu}}+\sum_{i,j,\nu\mu}t_{i,j}^{\nu,\mu}c_{i,{\nu}}^{\dagger}c_{j,{\mu}}\nonumber
\end{equation}
Here $c_{i,{\nu}}^{\dagger}(c_{i,{\nu}})$ is the creation(annihilation) operator of an orbital $\nu$ localised on an atom $i$. The terms signify the onsite orbital energies, intra-atomic orbital coupling, and inter-atomic orbital coupling respectively.\cite{klimeck2007atomistic} The summation of the last term is restricted to nearest neighbors only. $\epsilon$ and $t$ are treated as empirical fitting parameters for different materials and bond type, and they are expressed in terms of energy constants of $\sigma$ and $\pi$ bonds between the different atomic orbitals. In NEMO3D, the Keating model of the valence force field (VFF) method is applied to calculate the equilibrium position of atoms in the Ge/Si$_{1-x}$Ge$_x$ heterostructure.\cite{klimeck2002development} Contrasting to the continuum elastic theory based models, the resultant strain profile in NEMO3D captures the random alloy disorder, i.e. the effect of the random positioning of the Si atoms in the 'buffer' SiGe layers subjected to the inter-atomic potential. For a 10 nm wide Ge quantum well (QW) with 60 nm Si$_{0.2}$Ge$_{0.8}$ buffer simulated below and 10 nm Si$_{0.2}$Ge$_{0.8}$ above, a snippet of the VFF calculated atomic displacement field at equilibrium is shown in Fig.~\ref{fig:displacementprofile}. Importantly, due to interdiffusion,\cite{rodriguez2023linear}$ \varepsilon_{zz}$ relaxes to $0.5\%$ over the length of $\sim 3$ nm at the Si$_{0.2}$Ge$_{0.8}$/Ge interface. The fluctuation of $\varepsilon_{xy},\,\varepsilon_{xz},\,\varepsilon_{yz}$ results in non-zero shear strain gradient in the QW (Fig.~\ref{fig:strainZcuts2D}). For $\lvert B\rvert$=670 mT, the tight binding simulation of the hole EDSR Rabi frequency follows a different trend to that of the uniaxial strain model. In fact, Fig.~\ref{fig:fEDSRBdotEfuncplotNEMO} signifies the emergence of an asymmetric product relation: $f_\pi\propto \sin(\theta+\phi)$. The different optimum orientation of $f_\pi$ compared to the uniaxial $\mathbf{k}\cdot\mathbf{p}$ calculation indicates a different SOC mechanism for hole EDSR to the cubic Rashba or the orbital-B, originating from the random alloy disorder in realistic devices. The wave function obtained from TBA includes the effect of random alloy disorder, and notably, the maximum $f_\pi$ shows a $\sim$5x improvement in Fig.~\ref{fig:fEDSRBdotEfuncplotNEMO} over the uniaxial strain calculation (Figs.~\ref{fig:EDSR45nmnorough670mT1p5},\ref{fig:EDSR45nmnorough670mT}).

\subsubsection{Non-uniform strain due to gate electrodes}\label{sec:roughmodelmaps}

While the tight binding simulation takes into account the inhomogeneity of the quantum dot confinement as well as the short-range strain fluctuations, it does not incorporate effects stemming from the contraction of the metal gate-stack atop the heterostructure. 
Besides, large-scale tight binding simulations would incur a huge computational cost. In this work, a simpler and faster method based on the $\mathbf{k}\cdot\mathbf{p}$ model is adopted, where the atomistic strain due to the random alloy disorder and the gate-induced strain at cryogenic temperature are added as spatially varying profile $\varepsilon_{\alpha\beta}(x,y,z)$ in $H_\varepsilon$ (Eqn.~\ref{eq:3strainHam}). A single Ge dot is modeled after Ref.~\citenum{hendrickx2018gate} with the following gatestack parameters: 20 nm thick Al electrodes annealed atop the QW as source and drain to manipulate hole occupancy in the Ge layer, followed by a 17 nm layer of dielectric Al$_2$O$_3$. The final layer consists of Ti/Pd metal electrodes of thickness 5/35 nm, defining the quantum dot confinements. A qualitative sketch of the gatestack is given in Fig.~\ref{fig:GeomSideView}. Thermal strain arises at cryogenic experimental conditions due to the difference in contraction between the gatestack and the heterostructure; which is calculated using the finite element method (FEM) enabled in the thermal expansion and heat transfer modules of COMSOL Multiphysics with appropriate layered geometry and accurate material parameters. In Fig.~\ref{fig4:strainmaps} The thermal strain is superimposed on the strain due to random alloy disorder, highlighting the short-range and long-range fluctuations in the strain tensor components.\cite{corley2023nanoscale}

\subsection{Hole qubit operation in the presence of inhomogeneous strain}\label{sec:fullroughness}

\begin{figure*}[htbp!]
\subfloat[]{\begin{minipage}[c]{0.3\linewidth}
\vspace{0.1in}
\includegraphics[width=0.96\linewidth]{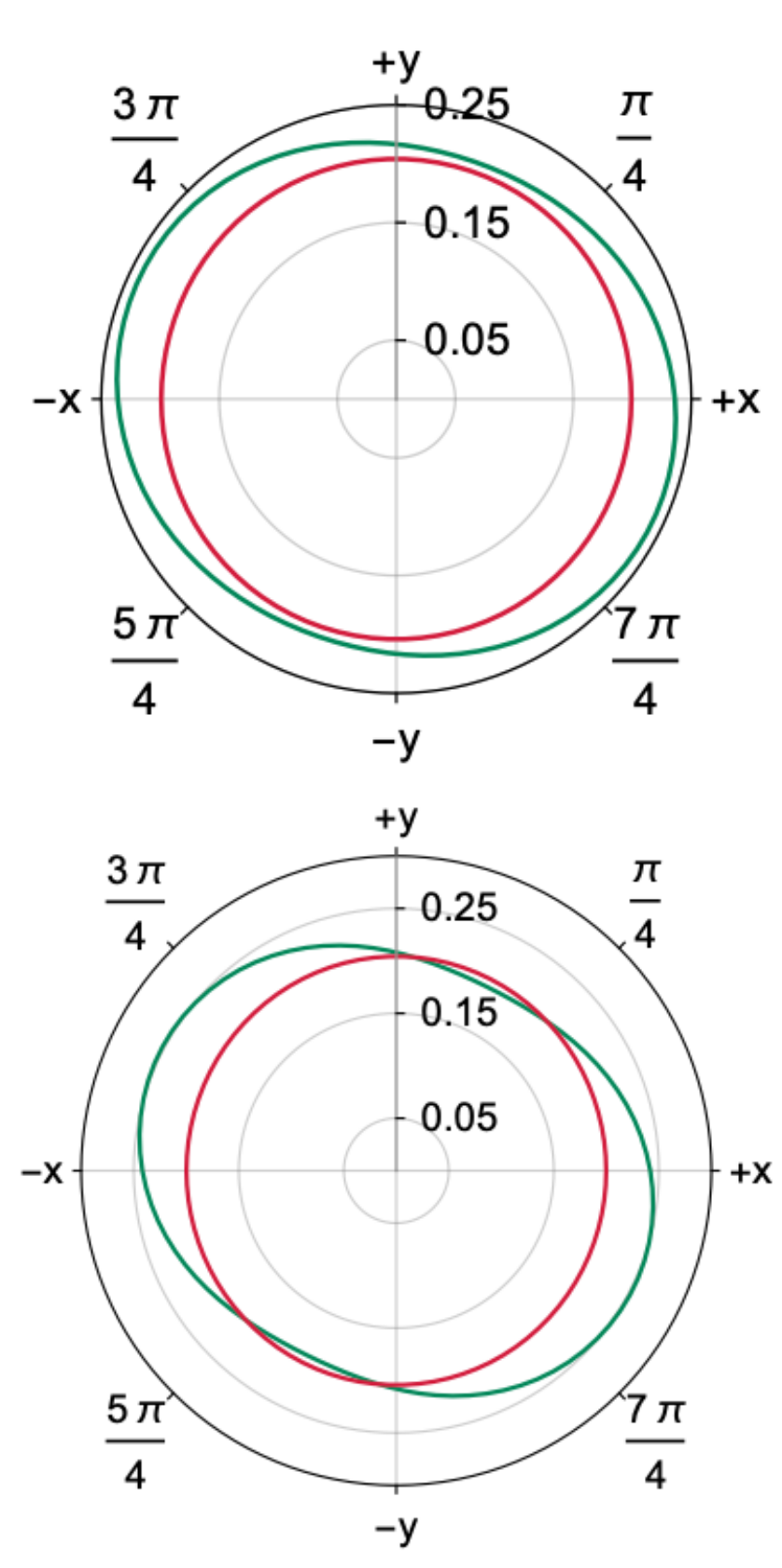}
\label{fig:gfactorSingleDot45Fin}
\end{minipage}}
\hfill
\subfloat[]{\begin{minipage}[c]{0.3\linewidth}
\includegraphics[width=\linewidth]{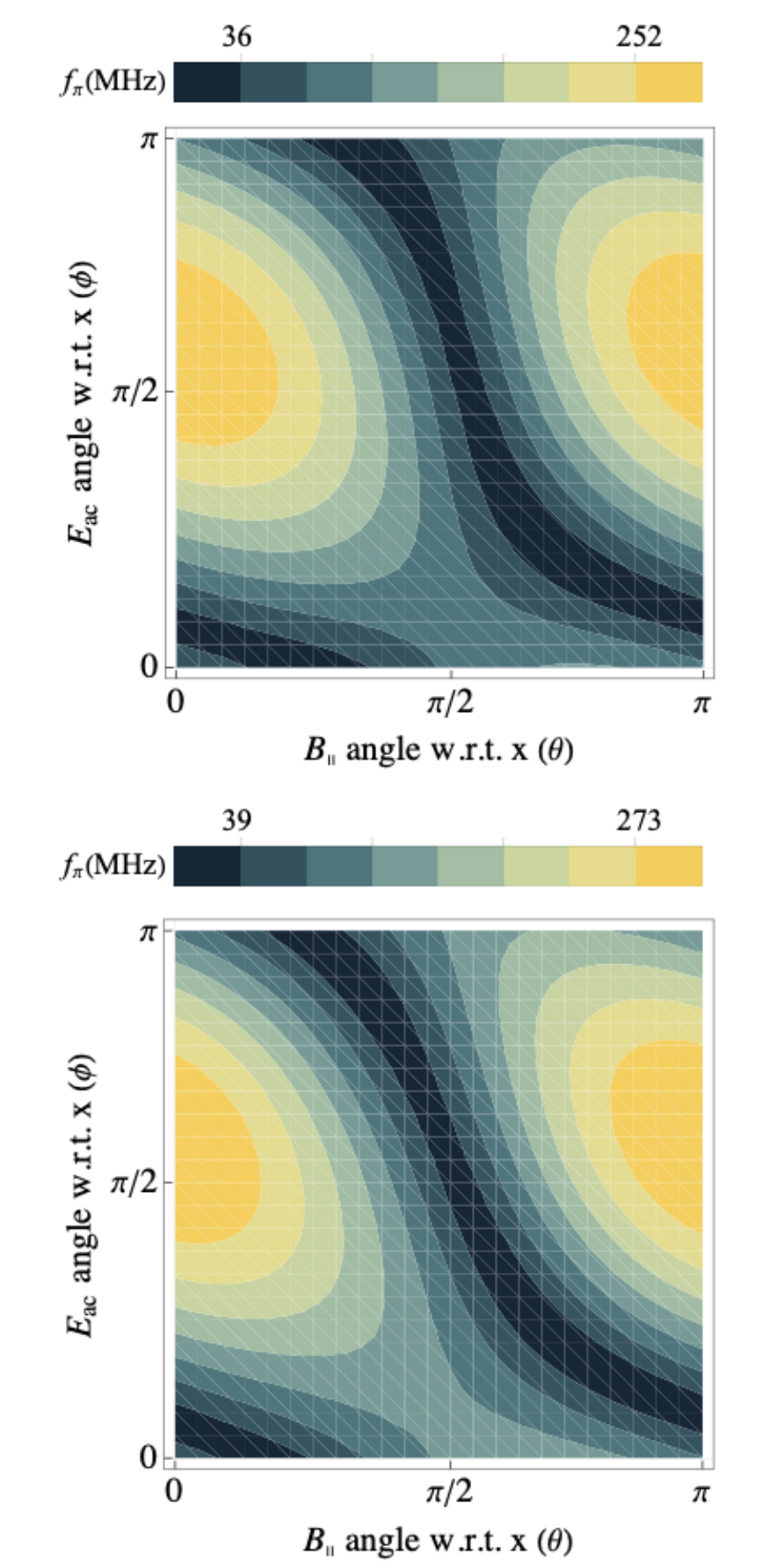}
\label{fig:EDSR45nmroughness}
\end{minipage}}
\hfill
\subfloat[]{\begin{minipage}[c]{0.4\linewidth}
\vspace{0.14in}
\includegraphics[width=0.96\linewidth]{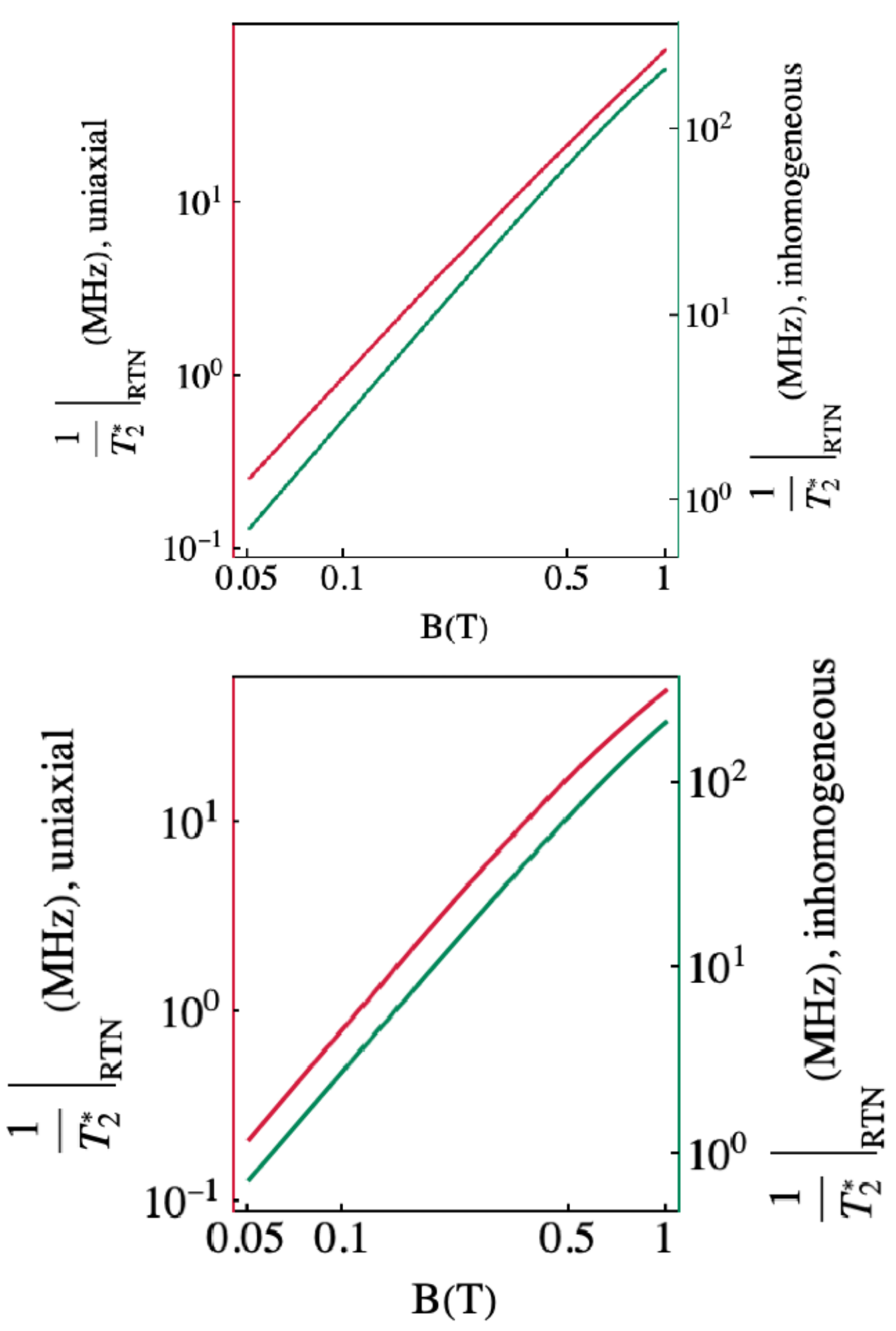}
\label{fig:QubitTimescales45nmT2star}
\end{minipage}}
\hfill
\\
\vspace{-0.25in}
\subfloat[]{\begin{minipage}[c]{0.5\linewidth}
\includegraphics[width=0.85\linewidth]{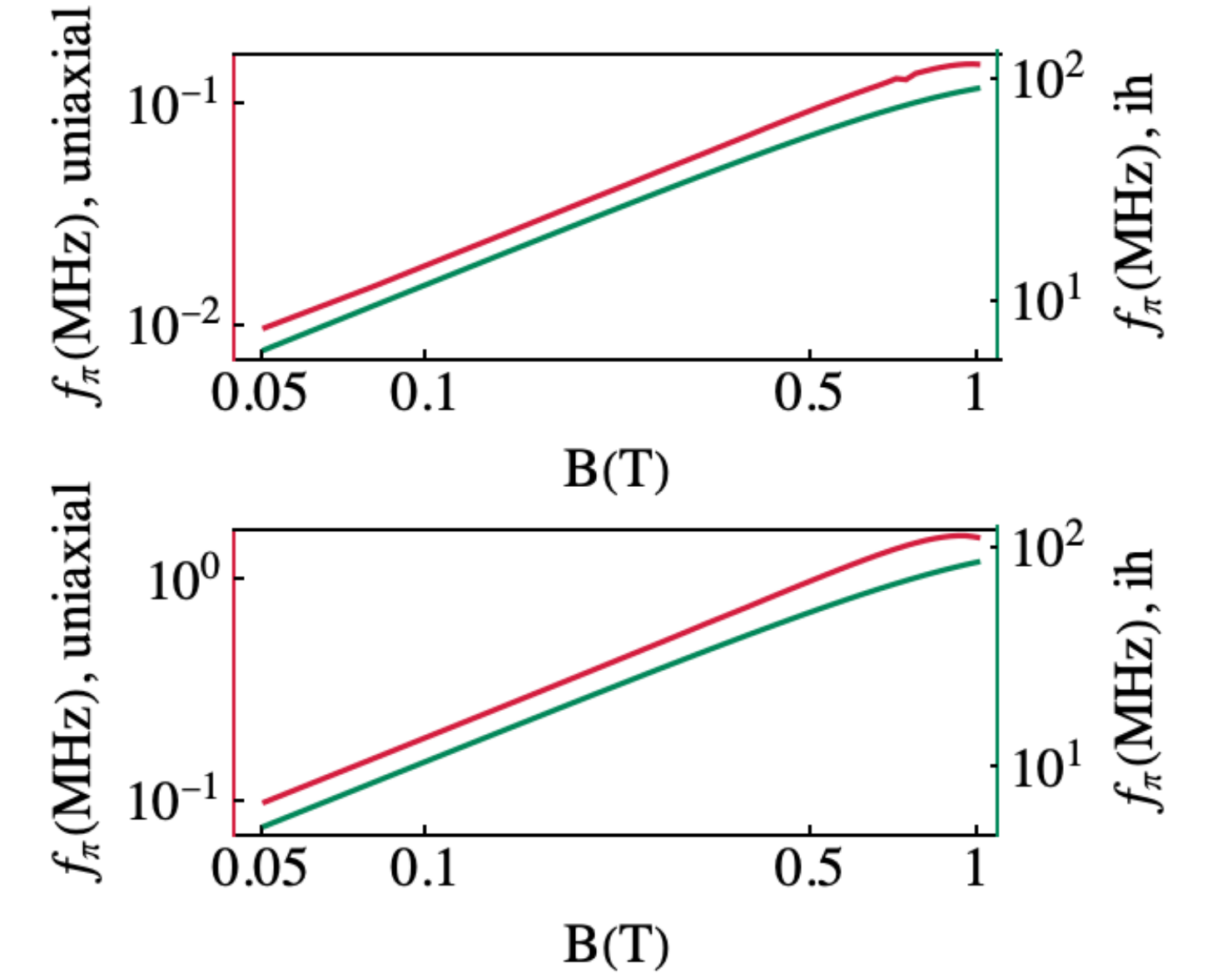}
\label{fig:QubitTimescales45nmTpi}
\end{minipage}}
\hfill
\subfloat[]{\begin{minipage}[c]{0.5\linewidth}
\includegraphics[width=0.85\linewidth]{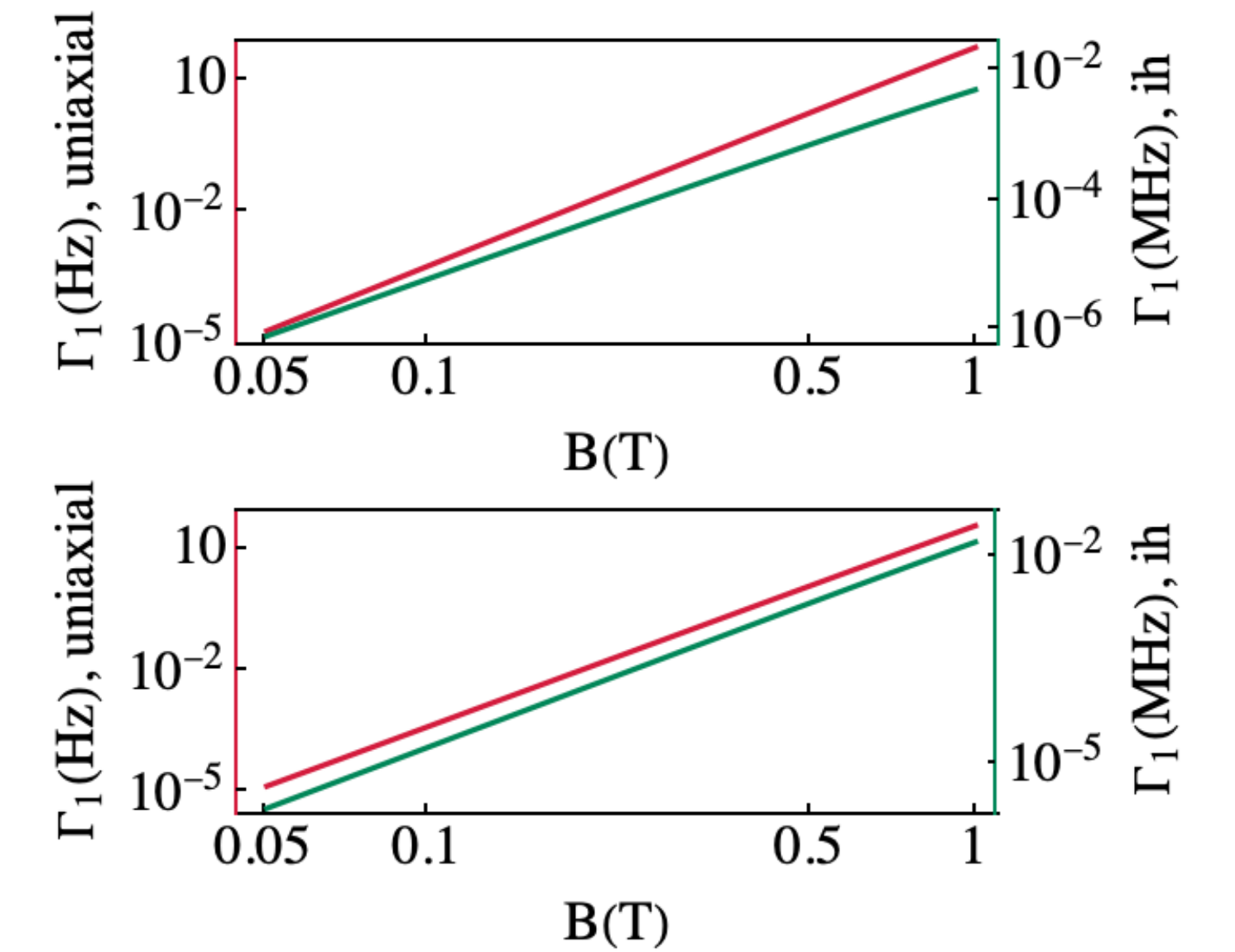}
\label{fig:QubitTimescales45nmT1}
\end{minipage}}
\hfill
\caption{{\bf Hole qubit properties with full strain profile.} For a circular dot of radius $a_d{=}45$ nm, a) The qubit $g$-factor is anisotropic and exhibits large $\sim \pi/4$ rotation of the principal axes. For all subfigures a-e), the top and bottom panels signify results for $F_z{=}1.5$ MV/m and $F_z{=}15$ MV/m, respectively. b) At $\lvert B\rvert{=}670$ mT, fast EDSR is predicted with gate time $\sim 10$ ns, and a similar $f_\pi(\theta,\phi)$ trend to the tight binding calculation is observed. Here, $|E_{\text{ac}}|{=}10$ kV/m. c) The $\propto \!\lvert B\rvert^2$ variation of the hole qubit dephasing rate ($(T_{2,RTN}^*)^{-1}$) due to the random telegraph noise from a single charge defect, with uniaxial strain (red) and inhomogeneous strain (green). The single charge defect is assumed to be situated at (50$\sqrt{2}$,50$\sqrt{2}$,0). d) The linear variation of the EDSR Rabi frequency $f_\pi$ w.r.t. the magnetic field strength $\lvert B\rvert$. In the presence of strain inhomogeneity (denoted by 'ih' on the green axes), $f_\pi$ is $\sim100\times$ higher than the uniaxial strain scenario (red). e) Relaxation rate ($\Gamma_1$) as a function of magnetic field strength.}
\label{fig5:roughnessqubit}
\end{figure*}

Complete knowledge of the inhomogeneous strain profile $\varepsilon_{\alpha\beta}(x,y,z)$ becomes important in the context of hole qubit properties mediated by spin-orbit coupling. The assumption of a uniaxial strain tensor only renormalizes the heavy hole-light hole gap in theory, but does not capture important phenomena in realistic devices: e.g. large anisotropy or site dependence of the $g$ tensor.\cite{john2024two,wang2024operating} Recently, Si and Ge hole qubits have reported improved control and fidelity in a multidot system by harnessing the strain inhomogeneity across the QDs and precise magnetic field orientation.\cite{liles2023singlet,hendrickx2024sweet,wang2024operating,rooney2025gate} With inhomogeneous strain, the EDSR Rabi frequency is calculated with the revised strain matrix elements of the $\mathbf{k}\cdot\mathbf{p}$ model: $\left\langle\Psi_l(x)\Psi_m(y)\Psi_n(z)\right|\varepsilon_{\alpha\beta}(x,y,z)\left|\Psi_{l'}(x)\Psi_{m'}(y)\Psi_{n'}(z)\right\rangle$. Following the description in Sec.~\ref{sec:3dkdotp}, in the $\{\Psi_l(x),\,\Psi_m(y),\,\Psi_n(z)\}$ basis with the simulation size $n,\,l,\,m\,{\in}[1,4]$, the resultant $H_\varepsilon$ Hamiltonian is a $256\times 256$ matrix. The modified eigenstates and eigenvalues of the total Hamiltonian in Eqn.~\ref{eq:2QDH} capture the effect of cumulative non-uniformity. The $f_\pi$ obtained lies in the range $\sim(30-300)$ MHz, with at least an order of magnitude improvement over the uniaxial assumption, and in much better agreement with experiments. The $f_\pi(\theta,\phi)$ variation is shown in Fig.~\ref{fig:EDSR45nmroughness}.

Planar hole quantum dots can be often described in the quasi-2D limit ($L_z{\ll}a_d$), since the large heavy hole-light hole gap implies that along with the top gate and orbital-B induced HH-LH matrix elements, inhomogeneous strain can also be treated perturbatively.
For a circular quantum dot, the hole EDSR Rabi frequency in the quasi-2D limit is evaluated as:
\begin{small}
\begin{equation}\label{eqn4:2DRabirate}
    f_\pi{=}\frac{\lvert B\rvert |E_{ac}|}{\Delta^2} \left(\alpha_{R3}f_1^{\theta,\phi}{+}\epsilon'_{xx}f_2^{\theta,\phi}{+}\epsilon'_{yy}f_3^{\theta,\phi}{+}\epsilon'_{xy}f_4^{\theta,\phi}+\epsilon'_{yx}f_5^{\theta,\phi}\right)
\end{equation}
\end{small}
In Eqn.~\ref{eqn4:2DRabirate}, the first term is the SIA Rashba, and strain inhomogeneity results in the next four terms, via both structural inversion asymmetric (SIA) linear-$k$ Rashba $\epsilon'_{xy}k_x\sigma_y{+}\epsilon'_{yx}k_y\sigma_x$ and bulk inversion asymmetric (BIA) Dresselhaus $\epsilon'_{xx}k_x\sigma_x{+}\epsilon'_{yy}k_y\sigma_y$ interactions.\cite{abadillo2023hole} Here $\epsilon'_{ij}$ consist of the strain gradients $\partial\varepsilon/\partial r$, and the angular functions $f_i^{\theta,\phi}$ describe the $f_\pi$ dependence on the respective magnetic field and microwave drive angles $\theta$ and $\phi$. Using Schrieffer-Wolff transformation, detailed expressions for $\alpha_{R3},\,\epsilon'_{ij}$ and the angular functions $f_i^{\theta,\phi}$ can be calculated (see supplementary). The subsequent $f_i^{\theta,\phi}$ plots suggest that the EDSR rate dependence on the $\mathbf{B}$ and $\Tilde{\mathbf{E}}_{\text{ac}}$ orientation boils down to the following relation:
\begin{eqnarray}\label{eq5:EDSRTrendEq}
    f_\pi\!&{=}&\!\frac{1}{\Delta^2}\left[\left(\alpha_{R3}+\epsilon'_{R}\right)\mathbf{B}\cdot\Tilde{\mathbf{E}}_{\text{ac}}{+}(\epsilon'_{D_{xy}}B_x\Tilde{E}_y{+}\epsilon'_{D_{yx}}B_y\Tilde{E}_x)\right]\nonumber\\
    &{=}&f_\pi^0\left[\cos(\theta-\phi)+r_1\cos\theta\sin\phi+r_2\sin\theta\cos\phi\right]
\end{eqnarray}
where, $f_\pi^0{=}(\alpha_R\lvert B\rvert |E_{ac}|)/\Delta^2$, with the accumulated Rashba coefficient $\alpha_R{=}\left(\alpha_{R3}+\epsilon'_{R}\right)$. Dimensionless ratios $r_1=\epsilon'_{D_{xy}}/\alpha_R$ and $r_2=\epsilon'_{D_{yx}}/\alpha_R$ denote the relative contribution of the Dresselhaus SOC w.r.t. the Rashba SOC. When strain inhomogeneity is absent, $r_1$, $r_2$ and $\epsilon'_{R}$ become zero, and $\alpha_{R}$ is maximum when $|\theta-\phi|{=}(n-1)\pi$; matching the result in Fig.~\ref{fig2:Uniaxial}. Note that fitting the plot in Fig.~\ref{fig:EDSR45nmnorough670mT} would require a perturbative analysis in higher order of $\mathbf{B}$ to account for the orbital effect.  

In summary, inhomogeneous strain induces the linear-$k$ Rashba contribution $\epsilon'_R$ as well the $\epsilon'_{D_{xy}}$ and $\epsilon'_{D_{yx}}$ linear-$k$ Dresselhaus contributions, with Dresselhaus terms ultimately dominating the EDSR Rabi frequency when a realistic strain profile is considered. The large $g$-factor anisotropy as well as rotation of the principal $g$-tensor is observed when strain nonuniformity is present (Fig.~\ref{fig:gfactorSingleDot45Fin}). Gate-induced thermal strain is responsible for the significant tuneability of the qubit $g$-tensor, making certain gate optimization possible.\cite{liles2021electrical} Fig.~\ref{fig:EDSR45nmroughness} demonstrates the strong presence of the Dresselhaus SOC at low as well as high top-gate fields. The variation of the qubit timescales with magnetic field strength is plotted next, with Fig.~\ref{fig:QubitTimescales45nmT2star} showing that inclusion of the full strain profile results in a higher RTN dephasing rate $(T_{2,RTN}^*)^{-1}$, with roughly a ${\propto}B^2$ dependence. The EDSR Rabi frequency varies linearly with the magnetic field strength (Fig.~\ref{fig:QubitTimescales45nmTpi}), and with strain inhomogeneity, $f_\pi$ improves by a factor $>10\times$ over uniaxial strain calculation. At the same time, the phonon-mediated relaxation rate $\Gamma_1$ increases nonlinearly with $\lvert B\rvert$ (Fig.~\ref{fig:QubitTimescales45nmT1}). 
For a more complete understanding of the hole spin qubit decoherence, in a future publication we will carry out an extended study of the effect of an ensemble of defects with the full strain profile, along the lines of Ref.~\citenum{wang2024dephasing}. With uniform uniaxial strain, the effect of the top-gate field on $f_\pi$ is significant.\cite{sarkar2023electrical} At an optimal $F_z{=}15$ MV/m the EDSR Rabi frequency is improved by an order of magnitude compared to that at $F_z{=}1.5$ MV/m (fig.~\ref{fig:QubitTimescales45nmTpi}), while the change in relaxation and RTN dephasing rates w.r.t. $F_z$ are smaller. However, in the presence of strain inhomogeneity the strain-induced SOC dominates, hence similar results are obtained for low and high top-gate fields for all of the timescales.

\section{Conclusion}\label{sec:conclu}

We have modeled a hole quantum dot in the Ge layer of a Ge/Si$_{1-x}$Ge$_x$ heterostructure using both the $\mathbf{k}\cdot\mathbf{p}$ and the tight binding frameworks independently, to analyze the effect of a realistic strain profile on the spectrum and operation of planar Ge hole qubits. We have identified that realistic strain inhomogeneity produces non-vanishing shear strain gradients across the Ge quantum dot, which leads to a $k$-linear Dresselhaus interaction, among other features. Our tight-binding calculation shows $\sim 5$x increase in the maximum EDSR Rabi frequency. In our $\mathbf{k}\cdot\mathbf{p}$ model, including both random alloy disorder-induced strain via valence force field calculation and metal gate-induced  strain via COMSOL, we demonstrate $>$10x improved hole EDSR Rabi frequency compared to uniaxial strain assumption, in better agreement with experimental observations. 

We have successfully accomplished our primary objective of integrating disorder and strain into the theoretical modeling of planar Ge hole qubits and gaining deeper insights into the resulting variability. Improving on these insights requires an extended benchmarking study over multiple architectures. Multiphysics calculations of planar Ge heterostructures with various buffer widths, SiGe compositions, or randomness would be vital in future studies of hole spin qubits. The method outlined in this work can be extended to encompass inter-dot variation of relevant hole spin qubit properties in multidot systems, while tight binding simulations can also offer important engineering insights.

\section{Methods}
Most of the results are obtained by theoretical analysis, following the effective mass approximation theory. For the perturbative analysis of electron dipole spin resonance (EDSR), as well as the two-qubit entanglement theory, we use Schrieffer-Wolff (SW) formalism as detailed in the supplementary material.

\section{Data Availability}
Data sharing not applicable to this article as no datasets were generated or analyzed during the current study.

\section{Acknowledgments} 
This project is supported by the Australian Research Council Center of Excellence in Future Low-Energy Electronics Technologies (project number CE170100039) and the Fellowship IL230100072.

\end{document}